\documentclass{article}

\usepackage{arxiv}

\usepackage[utf8]{inputenc} 
\usepackage[T1]{fontenc}    
\usepackage{hyperref}       
\usepackage{url}            
\usepackage{booktabs}       
\usepackage{amsfonts}       
\usepackage{nicefrac}       
\usepackage{microtype}      
\usepackage{lipsum}		
\usepackage{graphicx}
\usepackage{natbib}
\usepackage{doi}
\usepackage{amsmath}

\title{Profit Allocation in the We Media Value Chain: A Shapley Value-Based Approach}

\date{} 					

\author{\href{https://orcid.org/0009-0006-3794-2802}{\includegraphics[scale=0.06]{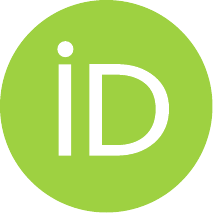}\hspace{1mm}Jianfei~Xu}\thanks{Funding Project: Youth Fund for Humanities and Social Sciences Research, Ministry of Education; Project Number: 15YJCZH232. Shanghai Art and Science Planning Project; Project Number: 2015C16. Shanghai University Young Teachers Training Program; Project Number: ZZGCD15018. Ideological and Political Research Project of Shanghai University of Engineering Science; Project Number: SZ201603.} \\
	Department of Management\\
	Shanghai University of Engineering Science\\
	Shanghai, China 201620 \\
	\texttt{mr.xujianfei@gmail.com} \\
        \And
	{\includegraphics[scale=0.06]{orcid.pdf}\hspace{1mm}Rui~Zhang\footnotemark[1]} \\
	Department of Management\\
	Shanghai University of Engineering Science\\
	Shanghai, China 201620 \\
	\texttt{rui5028369@126.com} \\
        \And
	{\includegraphics[scale=0.06]{orcid.pdf}\hspace{1mm}Junhui~Fan\footnotemark[1]} \\
	Department of Management\\
	Shanghai University of Engineering Science\\
	Shanghai, China 201620 \\
	\texttt{} \\
}

\renewcommand{\headeright}{}
\renewcommand{\undertitle}{}
\renewcommand{\shorttitle}{}

\hypersetup{
pdftitle={Profit Allocation in the We Media Value Chain: A Shapley Value-Based Approach},
pdfsubject={econ.TH},
pdfauthor={Jianfei Xu, Rui Zhang, Junhui Fan},
pdfkeywords={Shapley Value Method, AHP Hierarchical Analysis Method, Social Media Industry, Value Chain, Benefit Distribution},
}

\begin{document}
\maketitle

\begin{abstract}

The study takes the social media industry as its research subject and examines the impact of scientific innovation capabilities on profit distribution within the value chain of the social media industry. It proposes a specific solution to the profit distribution problem using an improved Shapley value method. Additionally, the AHP (Analytic Hierarchy Process) is employed to evaluate the profit distribution model, allowing the improved Shapley value method to better address the issue of profit allocation within the value chain of the social media industry. This approach ensures that each member receives a fair share of the profits, fostering strong cooperative relationships among members. Moreover, it compensates for the shortcomings of the traditional Shapley value method in addressing such problems to a certain extent.

\end{abstract}


\section{Introduction}

In recent years, with the continuous development of science and technology, the media industry has been undergoing profound changes. One specific manifestation of this transformation is the emergence of social media, which, as a new industry model, has been constantly challenging traditional media and reshaping various aspects of the media industry.

Social media has continuously evolved with advancements in information technology, and research on the subject has been growing both domestically and internationally. \cite{chen2015} found that domestic research primarily focuses on theoretical studies of social media, platform mechanisms, and specific social media objects. Other prominent areas of interest include public opinion regulation, crisis communication and management, and the relationship between social media and traditional media. Lastly, there is applied and functional research on new commercial, social, and legal phenomena brought about by the emergence of social media. \cite{yang2014} from the perspective of content marketing, explained the positive impact of content marketing on online interaction willingness using the Means-end Chains method. \cite{yu2013} identified two characteristics of online media business models: indirectness and complexity, and attempted to construct a flexible business model. \cite{ma2013} from a national perspective, argued that the era of social media places higher demands on fiscal democracy and suggested the active establishment of democratic fiscal supervision platforms to ensure the nation's democratic system keeps pace with the times. While the above scholars have conducted research on the development of the social media industry, most studies remain qualitative, focusing on areas such as business models, revenue models, and development models of social media. Few have quantitatively analyzed its value chain while considering the high-tech characteristics of social media. Unlike traditional media, social media’s innovative nature also encompasses scientific and technological innovation. Therefore, this paper analyzes the impact of technological innovation on the value chain of the social media industry and uses a Shapley value method based on AHP to analyze and model the profit distribution problem within the value chain of the social media industry.

\section{Influencing Factors}

With the rapid development of modern technology, scientific and technological innovation has become an independent factor contributing to production techniques. Innovation capability is primarily composed of technological autonomous innovation capability and technological cooperative innovation capability.

\begin{itemize}
    \item Technological autonomous innovation capability. This is determined by factors such as technological innovation investment capacity, design capability, production capability, high-tech industrialization capacity, and technological innovation output capability.
    \item Technological cooperative innovation capability. This mainly consists of elements like technological cooperation capacity and university-industry-research collaboration capacity \cite{leonard1992, lin2011, liu2007, chen2014}.
\end{itemize}

In the social media industry, technological autonomous innovation capability serves as the vitality of social media entities. By leveraging scientific and technological innovations, social media positions itself as a pioneer of its era. Therefore, social media must continuously rely on technological innovations to reshape communication methods, information acquisition processes, and more. Additionally, social media needs to innovate its content design to attract more customers and enhance resonance effects. Once a social media entity achieves innovative content design and a stable economic foundation, it must utilize its production capability to efficiently combine "blueprints" with real-world materials, producing social media products that can circulate on the internet. Furthermore, leveraging technological industrialization capacity to promote within the industry enhances the entity’s overall innovation output capacity, aiming for higher revenue. Apart from technological autonomous innovation capability, technological cooperative innovation capability is also crucial in influencing profit distribution in the value chain. Technological innovation is fundamentally based on theory, and social media entities must establish long-term strategic collaborations with academic institutions or other social media entities to secure timely theoretical support. By deepening technological cooperation capabilities and effectively integrating production with research, the economic benefits of scientific and technological innovation can be maximized.

\begin{figure}
    \centering
    \includegraphics[width=0.75\linewidth]{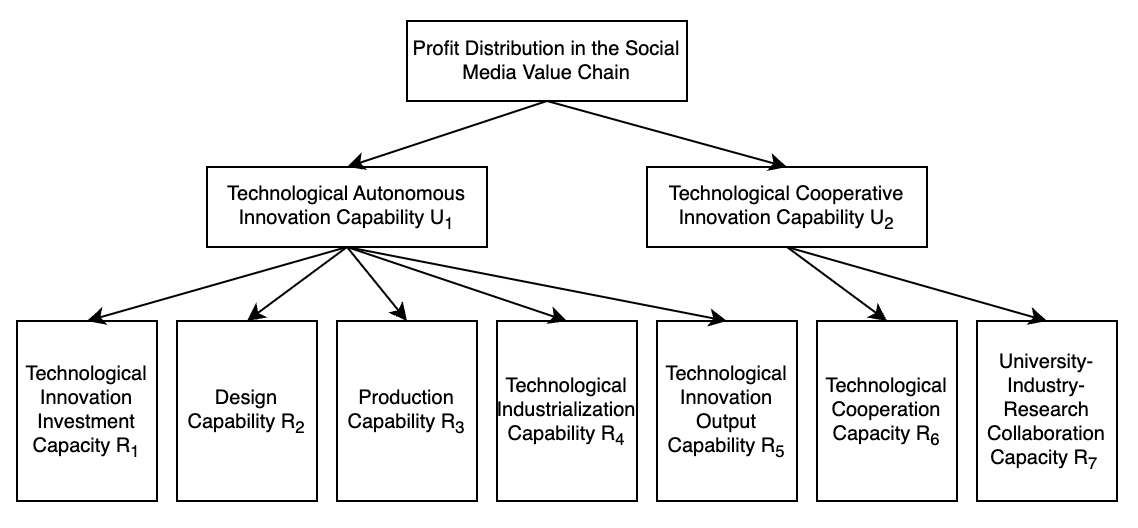}
    \caption{Indicators for Profit Distribution in the Social Media Value Chain}
    \label{fig:indicators_for_profit_distribution_in_the_social_media_value_chain}
\end{figure}

Therefore, this study evaluates the profit distribution in the value chain of the social media industry using the metrics of technological autonomous innovation capability and technological cooperative innovation capability. The analytic hierarchy process (AHP) is applied to subdivide these into seven sub-indicators, with the overall structure shown in Figure \ref{fig:indicators_for_profit_distribution_in_the_social_media_value_chain}.


\section{Establishment of the Profit Distribution Model for Social Media}

\subsection{Traditional Shapley Value Method}

Due to the fact that the modern social media industry often consists of entities that are interrelated in terms of benefits, the Shapley value principle is currently a relatively reasonable method for addressing the profit distribution within the social media industry. Let \( N \) represent the set of \( n \) members forming the coalition, \( S \) represent a coalition of members in the social media industry, and \( V(S) \) represent the characteristic function of coalition \( S \). For a member \( i \), the profit allocation result from collaboration is \( \phi_i(V) \). 

The profit distribution in the social media industry must satisfy the following conditions:  
1. Only members who contribute to the social media industry can receive a corresponding profit allocation.  
2. When a member participates in two collaborations simultaneously, the total distribution is the sum of the contributions from both collaborations.  
3. The collaboration must be efficient. Only when all effective profits are distributed reasonably within the social media industry can member \( n \) form a coalition in the social media industry \([9-10]\).

The Shapley value formula is therefore given as follows:  
\[
\phi_i(V) = \sum_{S \subseteq I} W(|S|) \left[ V(S \cup \{i\}) - V(S) \right], \quad i = 1, 2, 3, \ldots, n
\]
where:  
\[
W(|S|) = \frac{(n - |S|) ! \cdot (|S| - 1)!}{n!}
\]

In this formula:  
- \( W(|S|) \) is the weight coefficient of coalition \( S \) in the social media industry;  
- \( |S| \) is the number of elements in subset \( S \);  
- \( S \) represents all subsets of \( I \) that contain member \( i \).

\subsection{Improved Shapley Value Method}

The traditional Shapley value method assumes, as an initial condition for model construction, that all members have the same level of technological innovation capability. However, in real-world scenarios, this assumption is generally invalid, as the technological innovation capabilities provided by members of the social media industry vary to some extent. 

Therefore, to make the profit distribution in the value chain more reasonably reflect the individual value of members within the social media industry, and to maximize the motivation of each member to contribute to the industry, it is necessary to consider the impact of technological innovation capability on the overall profit of the social media industry and the profit distribution among its members.

Assume that the impact factor of technological innovation capability is \( G_i \), with the deviation defined as \( \Delta G_i = G_i - \frac{1}{n} \), where \( \sum_{i=1}^n G_i = 1 \) and \( \sum_{i=1}^n \Delta G_i = 0 \). Here, \( \Delta G_i \) represents the difference between the actual impact factor and the theoretical impact factor. Therefore, the profit allocation adjustment amount for member \( i \) in the social media industry is \( \Delta V(i) = V(S) \times \Delta G_i \), and the actual distributed profit becomes \( \Delta V'(i) = V(i) + \Delta V(i) \).

The improved Shapley value formula is therefore expressed as:
\[
\phi_i'(V) = \sum_{S \subseteq I} \frac{(n - |S|) ! \cdot (|S| - 1)!}{n!} \left\{ \left[V(S \cup \{i\}) - V(S) \right] + V(S) \left[G_i - \frac{1}{n} \right] \right\}, \quad i = 1, 2, 3, \ldots, n \tag{3}
\]

Here:
- If \( \Delta G_i \geq 0 \), it indicates that the member's actual contribution to the social media industry exceeds the theoretical contribution, and they should receive more profits than expected.  
- If \( \Delta G_i \leq 0 \), it indicates that the member's actual contribution is less than the theoretical contribution, and a portion of their profit should be deducted. The deduction amount is \( \Delta V(i) = V(S) \times |\Delta G_i| \), and the member's actual profit becomes \( \Delta V'(i) = V(i) - \Delta V(i) \).

From the above formula, it can be seen that the improved Shapley value method more accurately reflects the actual contribution of each member in the social media industry during the collaboration process. Moreover, it allows for a reasonable distribution of profits among the members of the social media industry based on actual circumstances.

\section{AHP Evaluation of Technological Innovation Capability in the Social Media Industry}

Due to the complexity of the actual profit distribution in the social media industry value chain compared to assumed scenarios, this paper employs the AHP method to calculate the weights of interdependent indicators in the profit distribution problem of the social media industry value chain. The overall hierarchical ranking is conducted to obtain the weights (\( W \)) of each indicator, followed by a consistency check of the results. The results are shown in Table \ref{fig:indicators_for_profit_distribution_in_the_social_media_value_chain}.

\begin{table}[htbp]
\centering
\begin{tabular}{lcc}
\toprule
\textbf{Factor ($R$)} & \textbf{Weight ($W$)} \\
\midrule
Technological Innovation Investment Capacity ($R_1$) & 0.4182 \\
Design Capability ($R_2$) & 0.2401 \\
Production Capability ($R_3$) & 0.1218 \\
Technological Industrialization Capability ($R_4$) & 0.1030 \\
Technological Innovation Output Capability ($R_5$) & 0.0442 \\
Technological Cooperation Capability ($R_6$) & 0.0351 \\
University-Industry-Research Collaboration Capability ($R_7$) & 0.0377 \\
\bottomrule
\end{tabular}
\caption{Weights of Each Impact Factor on the Profit Distribution in the Social Media Industry Value Chain}
\label{tab:weights_of_each_impact_factor_on_the_profit_distribution_in_the_social_media_industry_value_chain}
\end{table}

From the formula, the largest eigenvalue is calculated as \( \lambda_{\text{max}} = 7.5838 \), the consistency index is \( CI = 0.0973 \), and the consistency ratio is \( CR = \frac{CI}{RI} = 0.073 < 0.1 \), which satisfies the consistency check. The weights of each factor are calculated as follows:  
\[ 
[0.4182, 0.2401, 0.1218, 0.1030, 0.0442, 0.0351, 0.0377].
\]

Finally, based on calculations for each step, the influence factors of technological innovation capability on value chain profit distribution for three companies are obtained as follows:  
- Provider \( A \): \( G_A = 0.6648 \),
- Supplier \( B \): \( G_B = 0.2633 \),
- Supplier \( C \): \( G_C = 0.0703 \).

From an overall perspective, the evaluation results show that factors \( R_1, R_2, R_3, R_4, \) and \( R_5 \), all of which belong to the domain of technological autonomous innovation capability, have a greater influence on the profit distribution in the social media industry value chain. On the other hand, the factors with smaller weights, \( R_6 \) and \( R_7 \), fall under the domain of technological cooperative innovation capability. Both dimensions are indispensable and mutually complementary. Simultaneously, these results align with the current emphasis on the integration of technological layers into the social media industry [11], reflecting not only the decentralization of traditional dissemination pathways but also the unique value of technological innovation in social media.

From a local perspective, the evaluation results indicate that social media technological innovation emphasizes more on technological design capabilities. This aligns with its unique characteristic as a creative and artistic design industry. At the same time, attention must also be paid to improving the output ratio of technological innovation. By increasing the investment in technological innovation, social media industries can enhance their industrial technological capabilities.

\section{Case Analysis}

Assume a social media industry value chain consisting of three companies: \( A \), \( B \), and \( C \). In this value chain, company \( A \) serves as the provider, while companies \( B \) and \( C \) act as suppliers. By optimizing and integrating the core competencies of its member companies, the value chain aims to maximize overall profit.

If the three companies operate independently without collaboration, their profits are as follows:
\begin{itemize}
    \item Provider \( A \): 10 million yuan,
    \item Supplier \( B \): 5 million yuan,
    \item Supplier \( C \): 3 million yuan.
\end{itemize}

When companies collaborate in pairs, their profits are:
\begin{itemize}
    \item Collaboration between \( A \) and \( B \): 20 million yuan,
    \item Collaboration between \( A \) and \( C \): 15 million yuan,
    \item Collaboration between \( B \) and \( C \): 12 million yuan.
\end{itemize}

When all three companies \( A \), \( B \), and \( C \) collaborate simultaneously, their combined profit is 30 million yuan. Assume that the profit distributed to each company equals the total profit generated by the value chain. The assumed scenarios are shown in Table \ref{tab:profit_scenarios_for_different_collaboration_types}.

\begin{table}[htbp]
\centering
\begin{tabular}{lc}
\toprule
\textbf{Collaboration Type} & \textbf{Profit After Collaboration (million yuan)} \\
\midrule
\( A \) & 10 \\
\( B \) & 5 \\
\( C \) & 3 \\
\( A \cup B \) & 20 \\
\( A \cup C \) & 15 \\
\( B \cup C \) & 12 \\
\( A \cup B \cup C \) & 30 \\
\bottomrule
\end{tabular}
\caption{Profit Scenarios for Different Collaboration Types}
\label{tab:profit_scenarios_for_different_collaboration_types}
\end{table}

\section{Profit Distribution Calculation Using the Shapley Value Method}

Based on Table 2, this process shows that cooperation yields more profit than operating independently, with three-party collaboration generating more profit than two-party collaborations.

Firstly, according to the traditional Shapley value method, the profit distribution for each company \( \phi_A(V) \), \( \phi_B(V) \), and \( \phi_C(V) \) is calculated as follows:

\[
\phi_A(V) = \frac{(3-1)!(1-1)!}{3!} \times (1\,000 - 0) + \frac{(3-2)!(2-1)!}{3!} \times (2\,000 - 500) + 
\]
\[
\frac{(3-2)!(2-1)!}{3!} \times (1\,500 - 300) + \frac{(3-3)!(3-1)!}{3!} \times (3\,000 - 1\,200)
\]
\[
= 1\,383.3333
\]

\[
\phi_B(V) = \frac{(3-1)!(1-1)!}{3!} \times (500 - 0) + \frac{(3-2)!(2-1)!}{3!} \times (2\,000 - 1\,000) + 
\]
\[
\frac{(3-2)!(2-1)!}{3!} \times (1\,200 - 300) + \frac{(3-3)!(3-1)!}{3!} \times (3\,000 - 1\,500)
\]
\[
= 983.3333
\]

\[
\phi_C(V) = \frac{(3-1)!(1-1)!}{3!} \times (300 - 0) + \frac{(3-2)!(2-1)!}{3!} \times (1\,500 - 1\,000) + 
\]
\[
\frac{(3-2)!(2-1)!}{3!} \times (1\,200 - 500) + \frac{(3-3)!(3-1)!}{3!} \times (3\,000 - 2\,000)
\]
\[
= 633.3333
\]

Next, after evaluating the model using AHP, the influence factors of technological innovation capability on the value chain profit distribution are determined as:
\[
G_A = 0.6648, \quad G_B = 0.2633, \quad G_C = 0.0703.
\]
Substituting these influence factors into the improved Shapley value formula, the profits distributed to \( A \), \( B \), and \( C \) are calculated as follows:

\[
\phi'_A(V) = \frac{(3-1)!(1-1)!}{3!} \times \left[ 1\,000 + 1\,000 \times 0.6648 \right] + \frac{(3-2)!(2-1)!}{3!} \times \left[ 2\,000 - 500 + 2\,000 \times 0.6648 \right]
\]
\[
+ \frac{(3-2)!(2-1)!}{3!} \times \left[ 1\,500 - 300 + 1\,500 \times 0.6648 \right] + \frac{(3-3)!(3-1)!}{3!} \times \left[ 3\,000 - 1\,200 + 3\,000 \times 0.6648 \right]
\]
\[
= 2\,018.7083
\]

\[
\phi'_B(V) = \frac{(3-1)!(1-1)!}{3!} \times \left[ 500 + 500 \times 0.2633 \right] + \frac{(3-2)!(2-1)!}{3!} \times \left[ 2\,000 - 1\,000 + 2\,000 \times 0.2633 \right]
\]
\[
+ \frac{(3-2)!(2-1)!}{3!} \times \left[ 1\,200 - 300 + 1\,200 \times 0.2633 \right] + \frac{(3-3)!(3-1)!}{3!} \times \left[ 3\,000 - 1\,500 + 3\,000 \times 0.2633 \right]
\]
\[
= 1\,010.8333
\]

\[
\phi'_C(V) = \frac{(3-1)!(1-1)!}{3!} \times \left[ 300 + 300 \times 0.0703 \right] + \frac{(3-2)!(2-1)!}{3!} \times \left[ 1\,500 - 1\,000 + 1\,500 \times 0.0703 \right]
\]
\[
+ \frac{(3-2)!(2-1)!}{3!} \times \left[ 1\,200 - 500 + 1\,200 \times 0.0703 \right] + \frac{(3-3)!(3-1)!}{3!} \times \left[ 3\,000 - 2\,000 + 3\,000 \times 0.0703 \right]
\]
\[
= 751.6833
\]

Finally, through calculation, it can be concluded that the total profit obtained by Provider \( A \) under the given conditions is 2,018.7083 million yuan, the total profit obtained by Supplier \( B \) is 1,010.8333 million yuan, and the total profit obtained by Supplier \( C \) is 751.6833 million yuan. It can be observed that all three companies increased their profits through technological innovation. However, since \( A \) is the provider within the social media system and its technological innovation investment capability is stronger than that of \( B \) and \( C \), the returns from technological innovation for \( A \) are also greater than those for \( B \) and \( C \).

This demonstrates that such a profit distribution method is more reasonable and more likely to be accepted by the companies. On the one hand, this method enables companies to achieve higher profits within the value chain. On the other hand, it also encourages companies to continuously engage in technological innovation, thereby enhancing their technological innovation capabilities.

\section{Conclusion}

This paper uses the AHP method to quantitatively analyze the weight of scientific and technological innovation capabilities of social media members in the profit distribution of the value chain. By modifying the profit distribution method of the Shapley value, the distribution becomes more scientific and reasonable, making it easier to be accepted by members of the social media industry in practice. This ensures the sustainability and stability of the value chain.

\newpage
\bibliographystyle{unsrtnat}
\bibliography{references} 

\end{document}